\documentclass[runningheads,a4paper]{llncs}

\usepackage{amssymb}
\usepackage{graphicx}
\usepackage[T1]{fontenc}
\usepackage{amsmath}
\usepackage{mathbbol}
\usepackage[ruled]{algorithm2e}

\usepackage[T1]{fontenc}
\usepackage{amsmath}
\usepackage[ruled]{algorithm2e}
 
\usepackage{subfigure}
\usepackage{qcircuit}
\usepackage{algorithm2e}
\usepackage{dsfont}
\usepackage{amsfonts}
 \usepackage{times}
\usepackage{stmaryrd}
\usepackage{multirow}
\usepackage{tikz}
\usetikzlibrary{positioning}
 
\usepackage{url}

\urldef{\mailsa}\path|{Mingsheng.Ying}@uts.edu.au|    
\newcommand{\keywords}[1]{\par\addvspace\baselineskip
\noindent\keywordname\enspace\ignorespaces#1}

\def\>{\ensuremath{\rangle}}
\def\<{\ensuremath{\langle}}

\def\e{\ensuremath{\mathcal{E}}}

\newtheorem{thm}{Theorem}

\newtheorem{lem}{Lemma}
\newtheorem{defn}{Definition}

\newtheorem{exam}{Example}

\newcommand{\ket}[1]{|#1\rangle}

\newcommand{\tr}{{\rm tr}}
\newcommand {\spa } {{\rm span}}
\newcommand {\supp } {{\rm supp}}

\newcommand {\E } {{\mathcal{E}}}

\newcommand{\hs}{\mathcal{H}}

\begin{document}

\mainmatter

\title{Model Checking for Verification of Quantum Circuits}
\author{Mingsheng Ying}
\institute{Centre for Quantum Software and Information, University of Technology Sydney, Australia\\ 
 State Key Laboratory of Computer Science, Institute of Software, Chinese Academy of Sciences, China\\     
    Department of Computer Science and Technology, Tsinghua University, China\\
\email{Mingsheng.Ying@uts.edu.au}}
\titlerunning{Model Checking for Verification of Quantum Circuits}
\authorrunning{Ying}

\maketitle

\begin{abstract} In this talk, we will describe a framework for \textit{assertion-based verification} (ABV) of quantum circuits by applying \textit{model checking} techniques for quantum systems developed in our previous work, in which:\begin{itemize}
\item Noiseless and noisy quantum circuits are \textit{modelled} as operator- and super-operator-valued transition systems, respectively, both of which can be further represented by tensor networks. 
\item Quantum \textit{assertions} are specified by a temporal extension of Birkhoff-von Neumann quantum logic. Their semantics is defined based on the design decision: they will be used in verification of quantum circuits by simulation on classical computers or human reasoning rather than by quantum physics experiments (e.g. testing through measurements);    
\item \textit{Algorithms} for reachability analysis and model checking of quantum circuits are developed based on contraction of tensor networks. We observe that many optimisation techniques for computing relational products used in BDD-based model checking algorithms can be generalised for contracting tensor networks of quantum circuits. 
\end{itemize}
\end{abstract}

\section{Introduction}\label{Intro}

\textbf{\quad\ \ Assertion-based verification (ABV)} is a key methodology for functional verification of classical logic circuits and has been widely adopted in hardware industry. A major characteristic of ABV is that assertions are used for specifying design intent at a high level of abstraction and thus are ideal for using across multiple verification processes \cite{Foster}. An example application procedure of ABV was described in \cite{Marc} as follows: \begin{enumerate}\item A specification language such as PSL (Property Specification Language, IEEE 1850 standard) or SVA (SystemVerilog Assertions) is used to write the \textit{assertions} specifying the desired hardware properties. 
\item Verification is performed by formal methods or in a dynamic manner where a \textit{simulator} monitors the device under verification (DUV) and reports when and where assertions are violated. 
\item The information on assertion violation can be used in the \textit{debugging} process.\end{enumerate}

\textbf{Verification of quantum circuits} is emerging as an important issue as the rapid growth in the size of quantum computing hardware. Majority of the current research has been devoted to \textit{equivalence checking} of \textit{combinational} quantum circuits using various  quantum generalisations of BDDs (Binary Decision Diagrams), such as QuIDD \cite{Via04} \cite{Markov07}, QMDD \cite{Sei12} \cite{QMDD}. Recently, \textit{sequential} circuit model is emerging to play an important role in quantum computing and information processing; examples include quantum memories \cite{memories}, quantum feedback networks \cite{GJ08}, and RUS (Repeat-Until-Success) quantum circuits \cite{RUS}. A hardware description language was defined in \cite{QHDL} for specification of sequential quantum photonic circuits. An algorithm for equivalence checking of sequential quantum circuits is presented in \cite{WY18}. One can expect that as more and more sophisticated quantum hardware be physically realisable, more and more complicated verification problems will appear for quantum circuits, and assertion-based verification (ABV) will become an indispensable technology in future design automation for quantum computing (QDA).  

\textbf{Model checking quantum systems:} Essentially, assertion-based verification (ABV) of logic circuits can be seen as an important application of temporal logic and model checking. Research on extending model checking for quantum system has been conducted in the last fifteen years. Early work aimed at verification of quantum communication protocols \cite{Gay1} \cite{Lisbo} \cite{Gay2}. Targeting applications in analysis and verification of quantum programs \cite{Ying16}, several model checking techniques for quantum automata, quantum Markov chains and super-operator valued Markov chains have been developed in \cite{YL14} \cite{Concur14} \cite{Concur13} \cite{Feng13} (see \cite{YF21} for a more systematic exposition).    
However, a big gap between these quantum model checking techniques and their practical applications in verification of quantum circuits is still to be filled in. For near term applications, we believe that the following two considerations are crucial: 
\begin{itemize}\item \textit{Compact representations of quantum circuits}: As a compact representation, BDDs have played a key role in successful applications of model checking in verification of classical circuits \cite{Burch94}. 
As pointed out before, several quantum generalisations of BDDs have been employed in equivalence checking of quantum circuits. On the other hand, tensor networks - a mathematical tool successfully applied in simulation of quantum physical system for decades - have been widely used in simulation of large quantum circuits on classical computers in the last few years \cite{ETH17} \cite{Ali20} \cite{Riling} \cite{IBM19} \cite{Google19} . These representations should be helpful in implementing a more efficient model checker for quantum circuits.       
\item \textit{Simpler properties to be checked}: The previous research pursued theoretical generality and thus targeted checking general reachability and temporal logic properties of quantum systems. But a model checker (implemented on a classical computer) for such a purpose must be highly inefficient and only applicable to quantum circuits of very small sizes and depths. Thus, for realistic and in particular, near-term applications, we need to identify a class of simpler properties that can be efficiently checked by a current model checker for quantum systems.    
\end{itemize}

\textbf{In this talk,} we will describe a framework for \textit{assertion-based verification} (ABV) of quantum circuits by applying \textit{model checking} techniques for quantum systems developed in our previous work, in which quantum circuits are represented by tensor networks, and assertions about quantum circuits are specified using a simple temporal extension of Birkhoff-von Neumann quantum logic. The reason for using tensor network representation of quantum circuits is that algorithms for reachability analysis and model checking of quantum circuits can be conveniently implemented by contraction of tensor networks. More importantly, we observe that many optimisation techniques for computing relational products used in BDD-based model checking algorithms can be generalised for contracting tensor networks of quantum circuits. Our assertion language is chosen because it is actually useful for practical applications and at the same time, checking assertions written in it is much easier than others.      
 Furthermore, its semantics will be defined based on the design decision: the target application is verification of quantum circuits by simulation on classical computers (or human reasoning) rather than by quantum physics experiments (e.g. testing through measurements). We hope that focusing on this more realistic target, a model checker can be built for practical use in verification and debugging of near term quantum hardware. 

\section{Quantum Logic Circuits}

For convenience of the audience, let us start from a brief review on basics of quantum computing, with the emphasis on several basic models of quantum circuits. 

\subsection{Combinational Quantum Circuits}

Traditional combinational circuits are made from logic gates acting on wires. Combinational quantum circuits are quantum counterparts of them and made up of quantum (logic) gates, which are modelled by unitary operators. 

\textbf{Qubits (Quantum Bits):} The quantum counterpart of a bit is a qubit. A state of a single qubit is represented by a $2$-dimensional unit column vector $(\alpha, \beta)^T$, where $T$ stands for transpose, and complex numbers $\alpha, \beta$ satisfy the normalisation condition $\|\alpha\|^2+\|\beta\|^2=1$. It can be conveniently written in the Dirac's notation as $\ket{\psi}=\alpha \ket{0}+\beta \ket{1}$ with  $\ket{0}=(1,0)^T$, $\ket{1}=(0,1)^T$ corresponding to classical bits $0$ and $1$, respectively. 
Intuitively, this qubit is in a superposition of $0$ and $1$. In general, we use $q,q_1,q_2,...$ to denote qubit variables. Graphically, they can be thought of as wires in a quantum circuits. 
A state of $n$ qubits $q_1,...,q_n$ is then written as a $2^n$-dimensional unit complex vector $(\alpha_0,\alpha_1,...,\alpha_{2^n-1})^T$ or in the Dirac's notation: 
\begin{equation}\label{n-qubit}|\psi\rangle=\sum_{x\in\{0,1\}^n}\alpha_x |x\rangle=\sum_{x_1,...,x_n}\alpha_{x_1,...,x_n}|x_1,...,x_n\rangle\end{equation} where its norm $\||\psi\rangle\|=\sqrt{\sum_x|\alpha_x|^2}=1$, and we exchangeably use an $n$-bit string $x=x_1...x_n\in\{0,1\}^n$ and integer $x=\sum_{i=1}^{n}x_{i}\cdot 2^{i-1}.$ 

\textbf{Quantum Gates:} A gate on a single qubit is modelled by a $2\times 2$ complex matrix $U$. In general, a gate on $n$ qubits is described by a $2^n\times 2^n$ unitary matrix \begin{equation}\label{unitary}U=\left(U_{x,y}\right)_{x,y\in\{0,1\}^n}.\end{equation} The output of $U$ on an input $\ket{\psi}$ is quantum state $\ket{\psi^\prime}$. Its mathematical representation as a vector is obtained by standard matrix multiplication $|\psi^\prime\rangle=U\ket{\psi}$. To guarantee that $\ket{\psi^\prime}$ is always unit, $U$ must be unitary in the sense that $U^\dag U=I,$ where $U^\dag$ is the adjoint of $U$ obtained by transposing and then complex conjugating $U$. We often write $G\equiv U[q_1,...,q_n]$ for gate $U$ acting on qubits $q_1,...,q_n$.
 
\begin{exam}\label{ex-single}\begin{enumerate}\item The following are several frequently used single-qubit gates:\begin{enumerate} \item Hadamard gate: $H=\frac{1}{\sqrt{2}}\left(\begin{array}{cc}1 & 1\\ 1 &
-1\end{array}\right);$
\item The Pauli matrices:
$X=\left(\begin{array}{cc}0 & 1\\ 1 & 0\end{array}\right),\ \ Y=\left(\begin{array}{cc}0 & -i\\ i & 0\end{array}\right),\ \ Z=\left(\begin{array}{cc}1 & 0\\ 0 & -1\end{array}\right).$\end{enumerate}
\item Let $q_1,q_2$ be qubits. Then CNOT (controlled-$X$) gate $C[q_1,q_2]$ is a two-qubit gate with $q_1$ as the control qubit and $q_2$ as the target qubit and 
defined by the $4\times 4$ matrix $C=\left(\begin{array}{cc}I&0\\ 0&I\end{array}\right)$, where $I$ is the $2\times 2$ identity matrix. 
\end{enumerate}\end{exam}

\textbf{Combinational Quantum Circuits:} A combinational quantum circuit is a sequence of quantum gates: $C\equiv G_1...G_m$, where $m\geq 1$ and $G_1,...,G_m$ are quantum gates.
  
\begin{exam}The quantum circuit $Z[q_1]H[q_2]C[q_1,q_2]Y[q_1]H[q_2]$ consisting of five quantum gates is visualised in Figure \ref{combinational-quantum-circuit}.   
\begin{figure}
\centerline{
\Qcircuit @C=1em @R=0.9em {
q_1\ \ \ \ &\qw  & \gate{Z}  &\qw    &\ctrl{1}   &   \qw  & \gate{Y}  &\qw&\qw\\
q_2\ \ \ \ &\qw  & \gate{H}  &\qw    &\gate{X}   &   \qw  & \gate{H}  &\qw&\qw\\
}
}
    \caption{A combinational quantum circuit.}
    \label{combinational-quantum-circuit}
\end{figure}
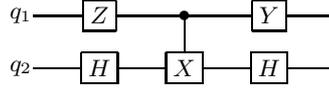
\end{exam}

\subsection{Noisy Quantum Circuits}

Fault-tolerant quantum computing is still out of the current technology's reach. To model noisy implementation of quantum circuits, we recall that a mixed state of an $n$-qubit system is an \textit{ensemble} $\{(|\psi_i\rangle, p_i)\}$ of its pure states, meaning that this system is in state $|\psi_i\rangle$ with probability $p_i$. Mathematically, this mixed state can be described by a $2^n\times 2^n$ matrix, called a density matrix, $\rho=\sum_ip_i|\psi_i\rangle\langle\psi_i|$, where $\langle\psi_i|$ is the conjugate transpose of $|\psi_i\rangle$ and thus a $2^n$-dimensional row vector. In particular, a pure state $|\psi\rangle$ can be identified with the outer product $|\psi\rangle\langle\psi|$. Then a noisy $n$-qubit gate can be modelled by a super-operator, often called a quantum channel in quantum information literature, which is a linear map $\E:\rho\rightarrow\E(\rho)$ from $2^n\times 2^n$ density matrices to themselves. A convenient way of representing $\E$ is the \textit{Kraus operator-sum form}: 
\begin{equation}\label{super-operator}\mathcal{E}(\rho)=\sum_{i}E_i\rho E_i^{\dag}\end{equation} for any density matrix $\rho$, where 
$\{E_i\}$ is a set of $2^n\times 2^n$ matrices  satisfying the normalisation condition $\sum_{i}E_i^{\dag}E_i= I_{2^n}$.
In particular, an idea $n$-qubit gate modelled by a unitary operator $U$ can be seen as a super-operator $\mathcal{U}:\rho\mapsto U\rho U^\dag$. 
\begin{exam}\label{flips} Several canonical noises on a single qubit are: 
\begin{enumerate}\item Bit flip: This noise flips the state of a qubit from $|0\rangle$ to $|1\rangle$ and vice versa with probability $1-p$, and is modelled by super-operator $\mathcal{N}_{\mathit{bf}}(\rho)=p\rho+(1-p)X\rho X$.  
\item Phase flip: This noise changes the phase of a qubit (that is, applies phase operator $Z$ on the qubit) with probability $1-p$, and is modelled by the super-operator $\mathcal{N}_{\mathit{pf}}(\rho)=p\rho+(1-p)Z\rho Z$.  
\item Bit-phase flip: This noise applies Pauli operator $Y$ on a qubit with probability $1-p$: $\mathcal{N}_{\mathit{bpf}}(\rho)=p\rho+(1-p)Y\rho Y$. Note that it is essentially a combination of a bit-flip and a phase flip because $Y=iXZ$.
 \end{enumerate}\end{exam}
 
 \subsection{Dynamic Quantum Circuits}

\ \ \ \ \ \textbf{Quantum Measurement:} The output of a combinational quantum circuit is a quantum state, which cannot be observed directly from the outside. To read out the outcome of computation, we have to perform a measurement at the end of the circuit. 
Mathematically, a quantum measurement on $n$ qubits is described by a family $M=\{M_m\}$ of $2^n\times 2^n$ matrices such that $\sum_mM_m^\dag M_m=I_{2^n}$, where $m$ denotes different possible outcomes. If one performs $M$ on the qubits   
in state $|\psi\rangle$, then outcome $m$ is obtained with probability $p_m=\|M_m|\psi\rangle\|^2$ and subsequently the state of these qubits will be changed to $\frac{M_m|\psi\rangle}{\sqrt{p_m}}$. 
More generally, if the $n$-qubit system is in a mixed state $\rho$, then outcome $m$ is obtained with probability $p_m=\tr(M_m^\dag M_m\rho)$ and its state will be changed to $\frac{M_m\rho M_m^\dag}{p_m}$.  
For example, the measurement in the computational basis is defined as $M=\{M_x:x\in\{0,1\}^n\}$ with $M_x=|x\rangle\langle x|$, and if it is performed on the qubits in a pure state (\ref{n-qubit}), then outcome $x\in\{0,1\}^n$ is obtained with probability $|\alpha_x|^2$ and subsequently the qubits will be in basis state $|x\rangle$.     

\textbf{Dynamic Quantum Circuits:} Quantum measurements are not only used for readout of the computational outcome at the end of a quantum circuit as described above. They may also occur at the middle of a quantum circuit where the measurement outcomes are used to conditionally control subsequent steps of the computation. This kind of circuits are called dynamic quantum circuits \cite{IBM} and several hardware platforms for quantum computing have matured to realise them. Formally, they can be are inductively defined as follows (see \cite{Ying16}, page 38):\begin{itemize}\item (Noiseless or noisy) quantum gates are dynamic quantum circuits;
\item If $C_1,C_2$ are dynamic quantum circuits, so is $C_1;C_2$; and \item If $M=\{M_m\}$ a measurement on qubits $q_1,...,q_n$, and for each possible outcome $m$, $C_m$ a dynamic quantum circuit , then  
$\mathbf{if}\ (\talloblong m\cdot M[q_1,...,q_n]
=m\rightarrow C_m)\ \mathbf{fi}$ 
is a dynamic quantum circuit. Intuitively, this conditional circuit performs measurement $M$ on qubits $q_1,...,q_n$, and then the subsequent computation is selected based on the measurement outcome: if the outcome is $m$, then the corresponding circuit $C_m$ follows.
\end{itemize}
Quantum teleportation is a simple example of dynamic quantum circuits. Another example is the dynamic circuit for quantum phase estimation shown as Figure 1 in \cite{IBM}.  
\begin{exam} Quantum teleportation is a protocol for transmitting quantum information (e.g. the exact state of an atom or photon) via only classical communication but with the help of previously shared quantum entanglement between the sender and receiver. It is one of the most surprising examples where entanglement helps to accomplish a certain task that is impossible in the classical world.
The quantum circuit teleporting a single qubit is shown in Figure \ref{fig:tele}. 
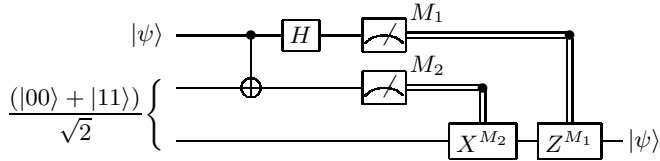
\begin{figure}		
		\centerline{
\Qcircuit @C=0.9em @R=0.8em {
	\lstick{\ket{\psi}} & \qw & \qw & \ctrl{1} & \gate{H} & \qw & \meter &  \raisebox{2em}{$M_1$} \cw & \cw & \control \cw \cwx[2] \\
		& \qw & \qw & \targ & \qw & \qw & \meter & \raisebox{2em}{$M_2$} \cw &\control \cw \cwx[1]  \\
	\lstick{\raisebox{3.5em}{$\displaystyle \frac{(\ket{00} + \ket{11})}{\sqrt{2}}\Biggl\lbrace$}} & \qw & \qw & \qw & \qw & \qw & \qw & \qw & \gate{X^{M_2}} & \gate{Z^{M_1}} & \qw & \ket{\psi}
}
}
\vspace{0.5em}
\caption{Quantum teleportation circuit}
\label{fig:tele}
\end{figure}

\end{exam}

\subsection{Sequential Quantum Circuits}

\begin{figure}
	    \centering
          \scalebox{0.7}{
	\Qcircuit @C=1em @R=.7em {
		\lstick{q_1} & \qw & \qw & \multigate{5}{\text{A combinational quantum circuit}} & \qw &  \qw \\
			 & & \lstick{\vdots} & & & \lstick{\vdots} \\
			\lstick{q_k} & \qw & \qw & \ghost{\text{A combinational quantum circuit}} & \qw & \qw \\
		\lstick{q_{k+1}} \qwx[6] & \qw & \qw > & \ghost{\text{A combinational   quantum circuit}} & \qw & \qw \qwx[6] \\
					 \lstick{\vdots} & & \lstick{\vdots} & & & \lstick{\vdots} \\
			 \lstick{q_{k+l}} & \qwx[2] & \qw > & \ghost{\text{A combinational quantum circuit}} & \qw \qwx[2] & & \\
		 	&  &  & &  & & & \\
			&  & \qw & \qw & \qw & & & \\ 
						&  &  &  &  & & & \\
			& \qw & \qw & \qw & \qw & \qw & & \\
		}
		 	}
	\vspace{0.5em}
		    \caption{A sequential quantum circuit}
		    \label{fig:seq}
		\end{figure}
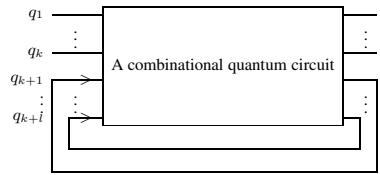

As is well-known, almost all practical digital devices contain (classical) sequential circuits. The output value of a combinational circuit is a function of only the current input value. In contrast, the output value of a sequential circuit depends on not only the external input value but also the stored internal information. All quantum circuits considered in the previous subsections are combinational. However, several recent applications appeal a \textit{sequential} model of quantum circuits, including quantum memories \cite{memories}, quantum feedback networks \cite{GJ08}, and RUS (Repeat-Until-Success) quantum circuits \cite{RUS}. A synchronous model of sequential quantum circuit was defined in \cite{WY18} and can be visualised as Figure \ref{fig:seq}, which looks similar to its classical counterpart, except:
\begin{itemize}\item The combinational part of a classical sequential circuit is modelled by a Boolean function; whereas the combinational part of a sequential quantum circuit is a unitary operator or a super-operator, depending on whether noise occurs in it. 
\item Certain measurements are needed at the end of qubits $q_1,...,q_k$ to readout classical information from their outputs.   
\end{itemize}

\subsection{Quantum Transition Systems as a Model of Quantum Circuits}\label{sec:transition}

A classical circuit can be conveniently described by a transition relation \cite{Burch94}. Similarly, 
all kinds of quantum circuits discussed above can be modelled by a quantum transition system defined in the following: 
 
\begin{defn}[Quantum Transition Systems] A quantum transition system (QTS) for a circuit with $n$ qubits consists of:\begin{enumerate}
 \item a finite set $L$ of locations, and an initial location $l_0\in L$; \item a set $\mathcal{T}$ of transitions: \begin{itemize}\item each transition $\tau\in \mathcal{T}$ is a triple $\tau=\langle l,l^\prime,\mathcal{E}\rangle$, often written as $\tau= l\stackrel{\mathcal{E}}{\rightarrow}l^\prime$ where $l,l^\prime\in L$ are the pre- and post-locations of $\tau$, respectively, and $\mathcal{E}$ is a super-operator on $2^n\times 2^n$ density matrices,\end{itemize} satisfying the normalisation condition:
$\sum\{|\tr[\mathcal{E}(\rho)]: l\stackrel{\mathcal{E}}{\rightarrow} l^\prime\in\mathcal{T}|\}=1$ 
for each $l\in L$ and $2^n\times 2^n$ density matrix $\rho$, where $\{|\cdot|\}$ stands for a multi-set, and trace $\tr(A)$ of a matrix $A$ is the sum of the entries on the diagonal of $A$. 
\end{enumerate}\end{defn}

In particular, for a noiseless quantum circuit, every transition $l\stackrel{\mathcal{E}}{\rightarrow} l^\prime$ is simply defined by a $2^n\times 2^n$ matrix $E$ such that $\mathcal{E}(\rho)=E\rho E^\dag$ for all density matrices $\rho$; for example, each quantum gate is defined as a unitary matrix $U$, and in a quantum measurement $M=\{M_m\}$, each branch corresponding to an outcome $m$ can be described by the measurement operator $M_m$.  

\begin{exam} The circuit of quantum teleportation in Figure \ref{fig:tele} can be modelled by the QTS in Figure \ref{fig:trans-tele}, where quantum operations are visualised by edges; for example, $\mathit{CX}_{1,2}$ on edge $l_0\rightarrow l_1$ denotes a CNOT on qubits $1$ and $2$, and $M_{2,1}$ on edge $l_2\rightarrow l_4$ means that a measurement is performed on qubit $2$ and outcome $1$ is obtained.   
\begin{figure}
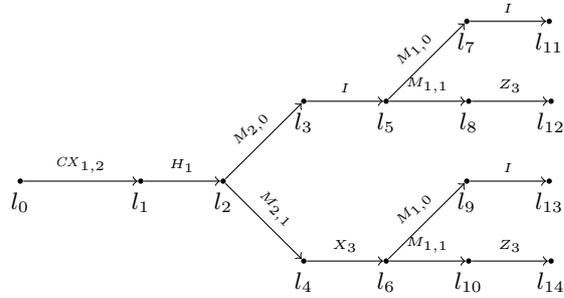

\centering
\tikz[location/.style={circle,minimum size=2pt,fill=black, inner sep=0pt}]{
	\node (l0) [location, label=-90:{$l_0$}] {};
	\node (l1) [location, right=1.5of l0, label=-90:{$l_1$}] {};
	\foreach \i/\j/\d in {2/1/right,3/2/above right,4/2/below right,6/4/right,5/3/right,7/5/above right,8/5/right,9/6/above right,10/6/right,11/7/right,12/8/right,13/9/right,14/10/right}{
		\node (l\i) [location, \d=of l\j, label=-90:{$l_{\i}$}] {};
	}
	\foreach \i/\j/\e in {0/1/\mathit{CX}_{1,2},1/2/H_1,2/3/M_{2,0},2/4/M_{2,1},3/5/I,4/6/X_3,5/7/M_{1,0},5/8/M_{1,1},6/10/M_{1,1},6/9/M_{1,0},7/11/I,8/12/Z_3,9/13/I,10/14/Z_3}{
		\draw[->] (l\i) --node[sloped,midway,above,font=\tiny] {$\e$} (l\j);
	}
}
	\vspace{0.5em}
\caption{Quantum teleportation circuit}
\label{fig:trans-tele}
\end{figure}
\end{exam}

\begin{remark}QTS's were first introduced in \cite{Gudder08} \cite{Feng13} where they are called quantum Markov chains. They were also used in defining invariants of quantum programs \cite{POPL'17}.  
\end{remark}
		
\section{Tensor Network Representation of Quantum Circuits}\label{sec:tensor}

In the last section, quantum circuits were defined in the traditional vector and matrix language of quantum mechanics. The shift from representing quantum circuits by matrices to tensor networks was proposed in \cite{IBM19} by identifying the following benefits in simulation of quantum circuits on classical computers: (i) quantum circuits can be arbitrarily partitioned into subcircuits; (ii) subcircuits can be simulated in arbitrary orders; and (iii) simulation results of subcircuits can be combined in arbitrary orders. From them, the reader might already notice that the advantage of tensor network representation of quantum circuits over matrices is very much similar to that of BDD representation of classical circuits over truth tables (i.e. Boolean matrices). In this section, we briefly review the basic idea of tensor networks and show how they can be used to represent quantum circuits.     

\subsection{Tensor Networks} 

A tensor is a multi-dimensional array of complex numbers.  We only consider a special class of tensors suitable for representing quantum circuits.
A tensor with an index set $\vec{q}=\{q_1,...,q_n\}$ is a mapping $T:\{0,1\}^{\vec{q}}\to\mathbb{C}$. We often write $T=T_{\vec{q}}$ or $T_{q_1,...,q_n}$ to indicate the indices. For two tensors $T_{\vec{p},\vec{r}}$ and $T_{\vec{q},\vec{r}}$ sharing indices $\vec{r}$, their \textit{contraction} is defined as a tensor $T_{\vec{p},\vec{q}}\stackrel{\triangle}{=}\mathit{Contract}(T_{\vec{p},\vec{r}}, T_{\vec{q},\vec{r}})$ by 
\begin{equation}\label{contract}T_{\vec{p},\vec{q}}(\vec{a},\vec{b})=\sum_{\vec{c}\in\{0,1\}^{\vec{r}}}T_{\vec{p},\vec{r}}(\vec{a},\vec{c})\cdot T_{\vec{q},\vec{r}}(\vec{b},\vec{c})\end{equation}
for any $\vec{a}\in\{0,1\}^{\vec{p}}$ and $\vec{b}\in\{0,1\}^{\vec{q}}$. 
Then a tensor network is a hyper-graph $H=(V,E)$, where a subset $E_0\subseteq E$ is chosen as open edges, and each vertex $v\in V$ is associated with a tensor of which the hyper-edges incident to $v$ are the indices. Thus, the hyper-edges between two vertices represent the indices shared by the two adjacent tensors. By contracting connected tensors in $H$, we can obtain a tensor $T_{H}$ with $E_0$ as its index set. It is easy to see that $T_{H}$ is independent of the order of contractions.   

\subsection{Representing Quantum States and Quantum Gates} The tensor representation of quantum states is straightforward. A pure state $|\psi\rangle$ of $n$ qubits $q_1,...,q_n$ given in Eq. (\ref{n-qubit}) can be represented by a tensor $T_{|\psi\rangle}\stackrel{\triangle}{=}T_{q_1,...,q_n}$ with 
$T_{q_1,...,q_n}(x_1,...,x_n)=\alpha_{x_1,...,x_n}$
for any $x_1,...,x_n\in\{0,1\}.$ Furthermore, a mixed state of $n$ qubits $q_1,...,q_n$ given as a $2^n\times 2^n$ density matrix $\rho=\left(\rho_{x,y}\right)_{x,y\in\{0,1\}^n}$ can be represented by a tensor $T_\rho\stackrel{\triangle}{=}T_{q_1,...,q_n,q_1^\prime,...,q_n^\prime}$ with \begin{equation}\label{tensor-density}T_{q_1,...,q_n,q_1^\prime,...,q_n^\prime}(x,y)=\rho_{x,y}\end{equation} for any $x,y\in\{0,1\}^n.$

Similar to the tensor representation (\ref{tensor-density}) of a density matrix, a (noiseless) quantum gate $U$ on $n$ qubits $q_1,...,q_n$ given as unitary matrix (\ref{unitary}) can be straightforwardly represented by a tensor $T_U\stackrel{\triangle}{=}T_{q_1,...,q_n,q_1^\prime,...,q_n^\prime}$ with 
$T_{q_1,...,q_n,q_1^\prime,...,q_n^\prime}(x,y)=U_{x,y}$ 
 for any $x,y\in\{0,1\}^n.$ 
To present a tensor representation of a noisy quantum gate $\mathcal{E}$ on $n$ qubits $q_1,...,q_n$, we assume that it is given in the Kraus representation (\ref{super-operator}), and define its matrix representation as \begin{equation}\label{matrix-represent}M_\mathcal{E}=\sum_i E_i\otimes E_i^\ast
\stackrel{\triangle}{=}\left(M_{x,y,x^\prime,y^\prime}\right)_{x,y,x^\prime,y^\prime\in\{0,1\}^n}\end{equation} where $E^\ast$ stands for the conjugate of $E$; that is, if $E=\left(E_{x,y}\right)$, then $E^\ast=\left(E_{x,y}^\ast\right)$, and $E_{x,y}^\ast$ is the conjugate of complex number $E_{x,y}$  for any $x,y\in\{0,1\}^n.$ Furthermore, if for each qubit $q_i$, we introduce a new copy $p_i$, then $M_\mathcal{E}$ can be represented by a tensor $T_\mathcal{E}\stackrel{\triangle}{=}T_{q_1,...,q_n,p_1,...,p_n,q_1^\prime,...,q_n^\prime,p_1^\prime,...,p_n^\prime}$ with
\begin{equation}\label{tensor-noise}T_{q_1,...,q_n,p_1,...,p_n,q_1^\prime,...,q_n^\prime,p_1^\prime,...,p_n^\prime}(x,y,x^\prime,y^\prime)=M_{x,y,x^\prime,y^\prime}\end{equation}
for any $x,y,x^\prime,y^\prime\in\{0,1\}^n.$

\subsection{Representing Quantum Circuits}

Now we can present a tensor network representation of quantum circuits by assembling the ingredients given in the previous subsections. Suppose we are given a combinational or sequential quantum circuit $C$ modelled as a quantum transition system. If we replace each (noiseless or noisy) gate in $C$ by its tensor representation, then we obtain a tensor network representation of $C$. Furthermore, one can compute its tensor $T_C$ by contraction (\ref{contract}). 
Moreover, if $|\psi\rangle$ or $\rho$ is an input to $C$, then the tensor representation of output $C|\psi\rangle$ or $C(\rho)$ can be computed as contraction $\mathit{Contract}(T_{|\psi\rangle},T_C)$ or $\mathit{Contract}(T_\rho,T_C)$, respectively.
When computing the tensor of a noisy quantum circuit, it is often more efficient to use contraction in combination with the following lemma, which gives a way for computing the matrix representations of the sequential and parallel compositions of noisy quantum gates.  
 
\begin{lem}\label{matrix-product}For any super-operators $\mathcal{E}$ and $\mathcal{F}$, we have: $$M_{\mathcal{E}\circ\mathcal{F}}=M_\mathcal{F} M_\mathcal{E};\qquad
M_{\mathcal{E}\otimes\mathcal{F}}=M_\mathcal{F}\otimes M_\mathcal{E}.$$\end{lem}

\subsection{Optimisations for Tensor Network Contraction}\label{optimisations}

It is obvious that computation required in the contraction of tensor networks of quantum circuits tends to be excessive as the growth of the number of qubits and the depth of circuits.    
In the last few years, many optimisation techniques have been proposed in the tensor network-based algorithms for simulation of quantum circuits on classical computers \cite{ETH17}, \cite{Ali20}, \cite{Riling}, \cite{IBM19}, \cite{Google19}. 
The main reason for employing tensor networks rather than large matrices in simulation of quantum circuits is that tensor networks can exploit the regularity and locality in the structure of quantum circuits. Essentially, the basic idea is similar to that of optimisation strategies in BDD-based algorithms (although this similarity has not been explicitly pointed out in the literature).
We believe that more BDD-optimisations can be adapted to computing tensor networks of quantum circuits, in particular when combined with their QST representations defined in Subsection \ref{sec:transition}; for example, Lemma \ref{matrix-product}   enables us to generalise    
 the partitioning technique in verification of classical circuits (see \cite{Burch94}, Section V) to the case of noisy quantum circuits.  
 For this purpose, we introduced a decision-diagram style data structure, called TDD (Tensor Decision Diagram), and showed that various operations of tensor networks essential in their applications can be conveniently implemented in TDDs \cite{Hong}.
 
\section{Reachability Analysis of Quantum Circuits}

We now start to consider the verification problem of quantum circuits. Many model checking problems about classical circuits (and other systems) can be reduced to a reachability problem. Reachability plays a similar role in model checking quantum systems \cite{YF21}. In this section, as a basis of verification techniques for quantum circuits, let us focus reachability of a simplest version of quantum transition systems, namely a \textit{quantum Markov chain} \cite{YY13}, which is defined as a pair $\mathcal{C}=\langle\mathcal{H},\e\rangle$, where $\mathcal{H}$ is a finite-dimensional Hilbert space as the system's state space, and $\e$ is a quantum operation (or super-operator) in $\mathcal{H}$ depicting transition of the system's state.
Roughly speaking, if the initial state is $\rho$, then the quantum Markov chain behaves as follows:
$\rho\rightarrow\e(\rho)\rightarrow \cdots\rightarrow\e^n(\rho)\rightarrow\e^{n+1}(\rho)\rightarrow\cdots.$
 
\subsection{Adjacency and Reachability}
   
As in the classical case, a graph structure is helpful for reachability analysis in quantum Markov chain $\mathcal{C}$. Let us first recall several notations needed in defining such a graph structure. For any $X\subseteq\mathcal{H}$, let $\spa (X)$ stand for the subspace spanned by $X$, i.e. the smallest subspace of $\mathcal{H}$ containing $Y$. The support $\supp (A)$ of an operator $A$ on $\mathcal{H}$ is the subspace spanned by the eigenvectors of $A$ associated with non-zero eigenvalues. For a family $\{X_i\}$ of subspaces of $\mathcal{H}_i$, we define their join as \begin{equation}\label{join}\bigvee_{i}X_i=\spa \left(\bigcup_i X_i\right).\end{equation} In particular, we write $X_1\vee X_2$ for the join of two subspaces $X_1$ and $X_2$. The image of a subspace $X$ of $\mathcal{H}$ under $\e$ is defined as $\e(X) = \bigvee_{|\psi\rangle\in X}\supp (\e(|\psi\rangle\langle\psi|))$, where $|\psi\rangle\langle\psi|$ is the density operator corresponding to pure state $|\psi\rangle$.
 
\begin{defn}[Adjacency Relation] Let $|\varphi\rangle, |\psi\rangle\in\mathcal{H}$ be pure states and $\rho, \sigma$ be mixed states (i.e. density matrices) in $\mathcal{H}$. Then
\begin{enumerate}
\item $|\varphi\rangle$ is adjacent to $|\psi\rangle$ in $\mathcal{C}$, written $|\psi\rangle\rightarrow|\varphi\rangle$, if $|\varphi\rangle\in\supp(\e(|\psi\rangle\langle\psi|))$. \item $|\varphi\rangle$ is adjacent to  $\rho$, written $\rho\rightarrow |\varphi\rangle$, if $|\varphi\rangle\in\e(\supp (\rho))$. \item $\sigma$ is adjacent to $\rho$, written $\rho\rightarrow\sigma$, if $\supp(\sigma)\subseteq\e(\supp (\rho))$.\end{enumerate}\end{defn}

Then as in classical graph theory, a path from a state $\rho$ to a state $\sigma$ in $\mathcal{C}$ is a sequence $\rho_0\rightarrow\rho_1\rightarrow\cdot\cdot\cdot\rightarrow\rho_n\ (n\geq 0)$ of adjacent states such that $\rho_0 = \rho$ and $\rho_n=\sigma$. For any two states $\rho$ and $\sigma$, if there is a path from $\rho$ to $\sigma$ then we say that $\sigma$ is reachable from $\rho$ in $\mathcal{C}$.

\begin{defn}[Reachable Subspace] For any state $\rho$ in $\mathcal{H}$, its reachable space in $\mathcal{C}$ is the 
subspace of $\mathcal{H}$ spanned by the states reachable from $\rho$:
$$\mathcal{R}_{\mathcal{C}}(\rho)=\spa\{|\psi\rangle\in\mathcal{H}:|\psi\rangle\ {\rm is\ reachable\ from}\ \rho\ {\rm in}\ \mathcal{C}\}.$$
\end{defn}

The following theorem from \cite{Concur12} gives a useful characterisation of reachable subspaces. It is essential a generalisation of Kleene closure in relational algebra. 

\begin{thm}\label{thm:reachablespace} Let $d=\dim \mathcal{H}$. Then for any state $\rho$ in $\mathcal{H}$, we have: 
\begin{equation}\label{transitive-1}
    \mathcal{R}_{\mathcal{C}}(\rho)=\bigvee_{i=0}^{d-1} \supp\left(\e^i(\rho)\right)=\supp\left(\sum_{i=0}^{d-1}\mathcal{E}^i(\rho)\right)
\end{equation} where $\e^i$ is the $i$th power of $\e$; that is, $\e^0=\mathcal{I}$ (the identity operation in $\mathcal{H}$) and $\e^{i+1}=\e\circ\e^i$ for $i\geq 0$.
\end{thm}

The reachable subspace $\mathcal{R}(\rho)$ can be viewed in a different way as the least fixed point of quantum predicate transformer (see \cite{YD10}, Section 8.4) $\mathcal{T}:\mathcal{S}(\mathcal{H})\rightarrow \mathcal{S}(\mathcal{H})$ defined by $\mathcal{T}(X)=\sup \rho \vee \E(X)$ for any $X\in \mathcal{S}(\mathcal{H})$.  

\subsection{Computing Reachable Subspaces}

Based on Theorem \ref{thm:reachablespace}, we can develop an algorithm for computing reachable subspaces in quantum Markov chain $\mathcal{C}$ using the tensor network representation of super-operator $\mathcal{E}$, with the help of the following:

\begin{lem}\label{matrix-compute} Let $|\Psi\rangle=\sum_k |kk\rangle$ be the (unnormalised) maximally entangled state in $\hs\otimes\hs$.  
 Then $(\mathcal{E}(A)\otimes I)|\Psi\rangle=M_\mathcal{E}(A\otimes I)|\Psi\rangle,$ where $I$ is the identity operator on $\hs$. 
 \end{lem}
 
The basic idea of the algorithm is as follows. Define state $|\eta\rangle= \sum_{i=0}^{d-1}\mathcal{E}^i(\rho)$ in $\hs$ and state $|\Phi\rangle=(\eta\otimes I)|\Psi\rangle$ in $\hs\otimes\hs$. Repeatedly using Lemma \ref{matrix-compute}, we obtain:
$$|\Phi\rangle= \sum_{i=0}^{d-1}\left(\mathcal{E}^i(\rho)\otimes I\right)|\Psi\rangle=\sum_{i=0}^{d-1}M_\E^i(\rho\otimes I)|\Psi\rangle.$$
Thus, state $|\Phi\rangle$ can be computed by contracting the tensor network representations of $M_\E$, $\rho$ and $|\Psi\rangle$. Finally, we can find the Schmidt decomposition of $|\Phi\rangle$:
$|\Phi\rangle=\sum_j p_j|j\rangle\otimes |j^\prime\rangle,$ where $p_j>0$ for all $j$. Then the reachable subspace $\mathcal{R}_\mathcal{C}(\rho)=\spa\{|j\rangle\}$ is computed. Of course, the optimisation techniques for contracting tensor networks discussed in Section \ref{optimisations} can be applied here and combined with Lemma \ref{matrix-product} when $\E$ comes from (sequential and parallel) compositions of smaller super-operators on subsystems.  
 
\section{Temporal Quantum Logic}

Now let us move on to consider the verification problem for a more general class of properties of quantum circuits. To specify these properties, we define an assertion language for quantum circuits in this section. We choose to simply use Birkhoff-von Neumann quantum logic \cite{BvN36} for specifying static behaviour of quantum circuits. 
To specify their behaviour over time, however, we need to introduce a temporal extension of Birkhoff-von Neumann logic. Several other temporal logics have been defined in the literature that are able to specify some sophisticated properties of quantum circuits than this logic. But we decide to adopt this simple temporal logic because its model checking can be much more efficiently implemented and may find practical applications in the early stage of quantum design automation.     

\subsection{Birkhoff-von Neumann Quantum Logic} Birkhoff-von Neumann logic is a \textit{propositional logic} for reason about (static properties of) quantum systems. We assume an alphabet consisting of: \begin{itemize}\item a set $\mathit{AP}$ of atomic propositions, ranged over by metavariables $X,X_1,X_2,...$; and \item propositional connectives $\neg$ (negation) and $\wedge$ (conjunction).\end{itemize}  
Given a Hilbert space $\mathcal{H}$ as the state space of the quantum circuit under consideration. We write $\mathcal{S}(\hs)$ for the set of its closed subspaces.
It is well-known that $(\mathcal{S}(\hs),\cap,\vee,\perp)$ is an orthomodular lattice with inclusion $\subseteq$ as its ordering, where $\cap, \vee$ and $\perp$ stand for intersection, join defined in Eq. (\ref{join}), and orthocomplement, i.e. $X^\perp=\{|\psi\rangle: |\psi\rangle\ {\rm is\ orthogonal\ to\ all}\ |\varphi\rangle\in X\}$. Then atomic propositions are interpreted as subspaces of $\hs$, i.e. elements of $\mathcal{S}(\hs)$, and connectives $\neg,\wedge$ are interpreted as $\perp$ and $\cap$, respectively.  
For each logical formula $A$, its semantics $\llbracket A\rrbracket$ is a subspace of $\hs$, meaning that the circuit's current state is within the region $\llbracket A\rrbracket$, and $\neg A$ indicates that the probability that the circuit's state enters the region $\llbracket A\rrbracket$ is zero. We can define $\vee$ (disjunction) by $A\vee B:=\neg(\neg A\wedge\neg B)$, and it is easy to see that $\llbracket A\vee B\rrbracket =\llbracket A\rrbracket \vee\llbracket B\rrbracket$ with the symbol $\vee$ in the right-hand side being join. Moreover, satisfaction of a proposition $A$ by a pure state $|\psi\rangle$ or a mixed state $\rho$ is simply defined as follows:
\begin{equation}\label{eq-atom}\varphi\models A\ {\rm iff}\ \varphi\in\llbracket A\rrbracket,\qquad\qquad \rho\models A\ {\rm iff}\ \supp(\rho)\subseteq \llbracket A\rrbracket.\end{equation}

\subsection{Computation Tree Quantum Logic}
A temporal extension of quantum logic can be naturally defined. For the limitation of space, we only consider computation tree quantum logic CTQL. Its syntax is the same as that of classical computation tree logic CTL:\begin{itemize}
\item State formulas:\qquad $\Phi::=A\ |\ \exists\varphi\ |\ \forall\varphi\ |\ \neg\Phi\ |\ \Phi_1\wedge\Phi_2$
\item Path formulas:\ \qquad $\varphi::=\ O\Phi\ |\ \Phi_1 U\Phi_2$\end{itemize}
except that $A$ stands here for a propositional formula in Birkhoff-von Neumann quantum logic rather than a classical (two-valued) proposition. 

\subsubsection{Simulation-Based Semantics:} We define a semantics of CTQL with the following \textit{design decision}: our verification of quantum circuits will be done by simulation on a classical computer. Therefore, no actual quantum measurement is performed for checking whether a quantum state $|\varphi\rangle$ or $\rho$ is in a subspace $X$, i.e. $|\varphi\rangle\models X$ or $\rho\models X$ according to Eq. (\ref{eq-atom}), and thus no quantum state decaying happens. 
Let $\mathcal{S}=\langle\mathcal{H},L,l_0,\mathcal{T}\rangle$ be a QTS. Then a configuration of $\mathcal{S}$ is a pair $(l,\rho)$, where $l\in L$ is a location and $\rho$ is a quantum state in $\hs$. We write $\mathcal{C}(\mathcal{S})$ for the set of configurations of $\mathcal{S}$. A sequence $\pi=(l_1,\rho_1)(l_2,\rho_2)$ $\cdots(l_{i-1},\rho_{i-1})(l_i,\rho_i)\cdots$ of configurations is a path in $\mathcal{S}$ if there exists a sequence $l_1\stackrel{\mathcal{E}_1}{\rightarrow}l_2\stackrel{\mathcal{E}_2}{\rightarrow}...\stackrel{\mathcal{E}_{i-1}}{\rightarrow} l_i\stackrel{\E_{i}}{\rightarrow}\cdots$ of transitions such that $\rho_{i+1}=\E_i(\rho_{i})$ for all $i$. We often write $\pi[i]=(l_{i+1},\rho_{i+11})$ for $i\geq 1$.
Then the satisfaction relation in CTL can be straightforwardly generalised to CTQL:\begin{defn}
\begin{enumerate}\item  Satisfaction $(l,\rho)\models\Phi$ for state formulas is defined as follows:   
\begin{enumerate}\item $(l,\rho)\models A$ iff $\supp(\rho)\subseteq\llbracket A\rrbracket$;
\item $(l,\rho)\models\exists\varphi$ iff $\pi\models\varphi$ for some path $\pi$ starting in $(l,\rho)$;
\item $(l,\rho)\models\forall\varphi$ iff $\pi\models\varphi$ for all paths $\pi$ starting in $(l,\rho)$;
\item $(l,\rho)\models\neg\Phi$ iff $\rho\not\models\Phi$;
\item $(l,\rho)\models\Phi_1\wedge\Phi_2$ iff $(l,\rho)\models\Phi_1$ and $(l,\rho)\models\Phi_2$.
\end{enumerate}\item  Satisfaction $\pi\models\varphi$ for path formulas is defined as follows: \begin{enumerate}\item $\pi\models O\Phi$ iff $\pi[1]\models\Phi$;
\item $\pi\models\Phi_1 U\Phi_2$ iff there exists $i\geq 0$ such that $\pi[i]\models\Phi_2$ and $\pi[j]\models\Phi_1$ for all $0\leq j<i$.
\end{enumerate}
\item We say that $\mathcal{S}$ with initial state $\rho$ satisfies $\Phi$, written $(\mathcal{S},\rho)\models\Phi,$ if $(l_0,\rho)\models\Phi$. 
\end{enumerate}\end{defn}

\begin{remark}  The above simulation-based semantics is fundamentally different from the measurement-based semantics of quantum temporal logics considered in the previous literature where the system's state is disturbed by a measurement, and the system's next step starts from the post-measurement state.\end{remark}

\section{Model Checking Quantum Circuits}

In this section, we show how model checking can be used in verification of the properties of quantum circuits specified in temporal logic CQTL introduced in the last section.   
\subsection{CTQL Model Checking}

Indeed, classical CTL model checking techniques can be adapted to solve the following:
\begin{itemize}\item \textbf{CTQL model checking problem}: Given a QTS $\mathcal{S}=\langle\mathcal{H},L,l_0,\mathcal{T}\rangle$, an initial state $\rho$ and a CTQL state formula $\Phi$. Check $(\mathcal{S},\rho)\models\Phi$?
\end{itemize}

The basic idea is to construct a classical transition system $\overline{\mathcal{S}}_\rho$ from a QTS with an initial state $\rho$ so that the above CTQL model checking problem is reduced to a CTL model checking problem. We construct $\overline{\mathcal{S}}_\rho=\langle\mathcal{C}(\mathcal{S})_\rho,\Rightarrow,(l_0,\rho),L\rangle$ as follows:\begin{itemize}\item Transition relation $\Rightarrow$ between configurations $(l,\rho),(l^\prime,\rho^\prime)\in\mathcal{C}(\mathcal{S})$ is defined by
\begin{equation}\label{cq-transition}(l,\rho)\Rightarrow (l^\prime,\rho^\prime)\ {\rm iff\ for\ some}\ \E:\ l\stackrel{\E}{\rightarrow}l^\prime\ {\rm and}\ \rho^\prime=\E(\rho);\end{equation}
\item We define $\mathcal{C}(\mathcal{S})_\rho$ as the set of configurations reachable from $(l_0,\rho)$ through $\Rightarrow$;
\item Configuration $(l_0,\rho)$ is defined as the initial state of $\overline{\mathcal{S}}_\rho$;
\item Propositional symbols $A$ in CTQL are interpreted as propositions in Birkhoff-von Neumann quantum logic and thus their semantics $\llbracket A\rrbracket$ are subspaces of $\hs$. However, in CTL for classical transition system $\overline{\mathcal{S}}_\rho$, they are considered as classical two-valued propositions, and labelling function $L$ interprets $A$ as follows: for each $(l,\sigma)\in\mathcal{C}(\mathcal{S})_\rho$, \begin{equation}\label{cq-prop}A\in L(l,\sigma),\ {\rm i.e.}\ (l,\sigma)\models A\ {\rm iff}\ \supp(\rho)\subseteq\llbracket A\rrbracket.\end{equation} 
\end{itemize}

The following simple lemma establishes a connection between CTQL for a QTS $\mathcal{S}$ and CTL for the classical transition system $\overline{\mathcal{S}}_\rho$ defined from $\mathcal{S}$ with an initial state $\rho$. 

\begin{lem}\label{lem-connect}For any CTQL state formula $\Phi$, any QTS $\mathcal{S}$ and any quantum state $\rho$ in $\mathcal{S}$, \begin{equation}\label{cq-equiv}(\mathcal{S},\rho)\models\Phi\ {\rm iff}\ \overline{\mathcal{S}}_\rho\models\Phi.\end{equation}\end{lem}

Note that in the left-hand side of (\ref{cq-equiv}), $\Phi$ is treated as a CTQL formula, but in the right-hand side, it is seen as a CTL formula in which atomic propositions $A$ are interpreted by labelling function $L$ defined in Eq. (\ref{cq-prop}).   

Based on Lemma \ref{lem-connect}, whenever $\mathcal{C}(\mathcal{S})_\rho$ is finite, then CTL model checking algorithms together with computations of (\ref{cq-transition}) and (\ref{cq-prop}) can be used to check whether $\overline{\mathcal{S}}_\rho\models\Phi$ or not.
However, it is possible that $\mathcal{C}(\mathcal{S})_\rho$ is infinite. In this case, we can apply bounded model checking to check the configurations reachable from $(l_0,\rho)$ through $\leq k$ steps.



 \subsection{Assertion-Based Verification of Quantum Circuits}
 
The above discussion indicates that assertions about quantum circuits written in CTQL can be verified by CTL model checking with some extra computations. It is well-known that a major practical hurdle in model checking applied to verifying classical circuits is the state space explosion problem. As one can imagine, this problem unavoidably occurs in the case of quantum circuits. The tensor network representation of quantum circuits discussed in Section \ref{sec:tensor}, together with various partitioning techniques for quantum transition systems defined in Section \ref{sec:transition} that exploit the locality in the circuits, can be a remedy to this issue. More explicitly, it is very helpful in computing reachable configurations $\mathcal{C}(\mathcal{S})_\rho$ and the labelling function (\ref{cq-prop}). The symbolic representation of quantum circuits using matrix-valued Boolean expressions proposed in \cite{YingJ} should also be useful.     

\section{Conclusion}

 In this talk, we presented a framework for assertion-based verification of quantum circuits by model checking with the help of tensor networks. The verified properties are \textit{qualitative} assertions written in a temporal extension of Birkhoff-von Neumann quantum logic. This modest aim of verifying only qualitative assertions is identified mainly for the reason that the verification algorithm can be more efficiently implemented and thus is actually useful in short-term practical applications. To check \textit{quantitative}  assertions (with probabilities) for quantum systems, some techniques have been developed in \cite{Feng13} \cite{YF21} \cite{Concur13} \cite{Yu}, but the involved computation will be overwhelming. To remedy this seemingly inevitable inefficiency of verifying  quantum circuits on classical computers, we are also trying to develop quantum algorithms for model checking quantum systems \cite{GWY}. 



\smallskip\

\textbf{Acknowledgment}: 
This work has been partly supported by the National Key R\&D Program of China (Grant No. 2018YFA0306701), the Australian Research Council (Grant No. DP210102449) and the National Natural Science Foundation of China (Grant No. 61832015).


\begin{thebibliography}{Lam94}

\bibitem{Lisbo} P. Baltazar, R. Chadha and P. Mateus, Quantum computation tree logic: model checking and
complete calculus, \textit{Int. J. of Quantum Information} 6(2008)219-36.

\bibitem{BvN36} G. Birkhoff and J. von Neumann, The logic of quantum mechanics, \textit{Annals of Mathematics} 37(1936)823-843.

\bibitem{RUS} A. Bocharov, M. Roetteler and K. M. Svore, Efficient synthesis of universal repeat-until-success quantum circuits, \textit{Physical Review Letters} 114(2015) art. no. 080502.

\bibitem{Marc} M. Boule, J. -S. Chenard and Z. Zilic, Assertion checkers in verification, silicon debug and in-field diagnosis, In: \textit{Proc. of the 8th IEEE Int. Symp. on Quality Electronic Design}, 2007, pp. 613-620. 

\bibitem{Burch94} J. R. Burch, E. D. Clarke, D. E. Long, K. L. McMillan and D. L. Dill, Symbolic model checking for sequential circuit verification, \textit{IIEEE Trans. Comput. Aided Des. Integr. Circuits Syst} 13(1994)401-424. 

\bibitem{QMDD} L. Burgholzer and R. Wille, Advanced equivalence checking for quantum circuits, \textit{IEEE Trans. Comput. Aided Des. Integr. Circuits Syst.}, 2021. 


\bibitem{IBM} A. D. C\/{o}rcoles, M. Takita, K. Inoue, S. Lekuch, Z. K. Minev, J. M. Chow and J. M. Gambetta, Exploiting dynamic quantum circuits in a quantum algorithm with superconducting qubits, \textit{arXiv}: 2102:01682. 

\bibitem{Gay2} T. A. Davidson, S. J. Gay, H. Mlnarik, R. Nagarajan and N. Papanikolaou, Model checking for communicating quantum processes, \textit{Int. J. of Unconventional Computing} 8(2011)73-98.

\bibitem{Feng13} Y. Feng, N. K. Yu and M. S. Ying, Model checking quantum Markov chains, \textit{J. of Computer and System Sciences}, 79(2013)1181-1198.

\bibitem{Foster} H. Foster and E. Marschner, Assertion-based verification, In: L. Lavagno, G. Martin,   
I. L. Markov, L. K. Scheffer (eds.), \textit{Electronic Design Automation for IC System Design, Verification, and Testing}, CRC Press, 2016, pp. 441-460. 

\bibitem{Gay1} S. J. Gay, R. Nagarajan and N. Papanikolaou, QMC: a model checker for quantum systems, In: \textit{Proc. of CAV'2008}, pp. 543-47.

\bibitem{GJ08} J. E. Gough and M. R. James, Quantum feedback network: Hamiltonian formulation, \textit{Communications in Mathematical Physics} 287(2008)1109-1132. 

\bibitem{GWY} J. Guan, Q. S. Wang and M. S. Ying, An HHL-based algorithm for computing hitting probabilities of quantum random walks, \textit{Quantum Information \& Computation} 2021. 

\bibitem{Gudder08} S. Gudder, Quantum Markov chains, {\em J. of Mathematical Physics}, 49(2008) art. no. 072105.

\bibitem{ETH17} T. H\"{a}ner and D. S. Steiger, 0.5 petabyte simulation of a 45-qubit quantum circuit, In: \textit{Proc. of SC'2017}, pp. 1-10.

\bibitem{Hong} X. Hong, X. Z. Zhou, S. J. Li, Y. Feng and M. S. Ying,
A tensor network based decision diagram for representation of quantum circuits, \textit{arXiv}: 2009.02618. 

\bibitem{Ali20} C. J. Huang, F. Zhang, M. Newman, J. J. Cai, X. Gao, Z. X. Tian, J. Y. Wu, H. H. Xu, H. J. Yu, B. Yuan, M. Szegedy, Y. Y. Shi and J. X. Chen, Classical simulation of quantum supremacy circuits, 
\textit{arXiv}:2005.06787. 

\bibitem{memories} J. Kerckhoff, H. I. Nurdin, D. S. Pavlichin and H. Mabuchi, Designing quantum memories with embedded control: photonic circuits for autonomous quantum error correction, \textit{Physical Review Letters} 105(2010) art. no. 040502. 


\bibitem{Riling} R. L. Li, B. J. Wu, M. S. Ying, X. M. Sun and G. W. Yang, Quantum supremacy circuit simulation on Sunway TaihuLight, \textit{IEEE Trans. on Parallel and Distributed Systems} 31(2020)805-816.

\bibitem{Concur14} Y. J. Li and M. S. Ying. (Un)decidable problems about reachability of quantum systems, In:
\textit{Proc. of CONCUR'2014}, pp. 482-496.

\bibitem{IBM19} E. Pednault, J. A Gunnels, G. Nannicini, L. Horesh, T. Magerlein, E. Solomonik, E. W. Draeger, E. T. Holland and R. Wisnieff, Breaking the 49-qubit barrier in the simulation of quantum circuits, \textit{arXiv}:1710.05867.

\bibitem{Sei12} J. Seiter, M. Soeken, R. Wille and R. Drechsler, Property checking of quantum circuits using quantum multiple-valued decision diagrams, In: \textit{International Workshop on Reversible Computation
(RC)}, 2012, pp. 183-196.

\bibitem{QHDL} N. Tezak, A. Niederberger, D. S. Pavlichin, G. Sarma and H. Mabuchi, Specification of photonic circuits using quantum hardware description language, \textit{Philosophical Transactions of the Royal Society A} 370(2012)5270-5290.  

\bibitem{Via04} G. F. Viamontes, I. L. Markov and J. P. Hayes, Improving gate-level simulation of quantum circuits, \textit{Quantum Information Processing} 2(2004)347-379. 

\bibitem{Markov07} G. F. Viamontes, I. L. Markov and J. P. Hayes, Checking equivalence of quantum circuits and states, In: \textit{Proc. ICCAD'2007}, pp. 69-74.

\bibitem{Google19} B. Villalonga, S. Boixo, B. Nelson, C. Henze, E. Rieffel, R. Biswas and S. Mandra, A flexible high-performance simulator for verifying and benchmarking quantum circuits implemented on real hardware, \textit{NPJ Quantum Information}  5(2019)art. no. 86.
 
\bibitem{WY18} Q. S. Wang and M. S. Ying, Equivalence checking of sequential quantum circuits, \textit{arXiv}:1811.07722.

\bibitem{Ying16} M. S. Ying, \textit{Foundations of Quantum Programming}, Morgan Kaufmann, 2016.

\bibitem{YD10} M. S. Ying, R. Y. Duan, Y. Feng and Z. F. Ji, Predicate transformer semantics of quantum programs, In: I. Mackie and S. Gay (eds.), \textit{Semantic Techniques in Quantum Computation}, Cambridge University Press 2010, pp. 311-360.

\bibitem{YF21} M. S. Ying and Y. Feng, \textit{Model Checking Quantum Systems: Principles and Algorithms}, Cambridge University Press, 2021. 

\bibitem{YingJ} M. S. Ying and Z. F. Ji, Symbolic verification of quantum circuits, \textit{arXiv}: 2010.03032. 

\bibitem{YL14} M. S. Ying, Y. J. Li, N. K. Yu and Y. Feng, Model-checking linear-time properties of quantum systems, \textit{ACM Transactions on Computational Logic} 15(2014) art. no. 22.

\bibitem{POPL'17} M. S. Ying, S. G. Ying and X. D. Wu, Invariants of quantum programs: characterisations and generation, In: \textit{POPL} 2017, pp. 818-832.  

\bibitem{YY13} M. S. Ying, N. K. Yu, Y. Feng and R. Y. Duan, 
Verification of quantum programs, \textit{Science of Computer Programming} 78(2013)1679-1700.

\bibitem{Concur13} S. G. Ying, Y. Feng, N. K. Yu and M. S. Ying. Reachability probabilities of quantum Markov chains, In: \textit{Proc. of CONCUR'2013}, pp. 334-348.

\bibitem{Concur12} N. K. Yu and M. S. Ying, Reachability and termination analysis of concurrent quantum programs, In: \textit{Proc. of  CONCUR'2012}, pp. 69-83.

\bibitem{Yu} N. K. Yu, Quantum temporal logic, \textit{arXiv}:1908.00158.

\end{thebibliography}
 \end{document}